\documentclass[letter]{aa} 

\usepackage{graphics}
\usepackage{longtable}
\usepackage{txfonts}
\usepackage{graphicx}
\usepackage{hyperref}

\begin{document}

\title{$\delta$ Cen: a new binary Be star detected by VLTI/AMBER spectro-interferometry \thanks{Based on observations made with ESO telecopes at La Silla Paranal Observatory under GTO programme IDs 080.D-0131(A) and 080.C-0099(B)}}

\authorrunning{A. Meilland et al.}
\titlerunning{$\delta$ Cen: a new binary Be star detected by VLTI/AMBER}

   \author{A. Meilland \inst{1},  F. Millour  \inst{1}, Ph. Stee \inst{2}, A. Spang \inst{2}, R. Petrov\inst{2}, D. Bonneau \inst{2}, K. Perraut \inst{3}, and  F. Massi\inst{4}}

   \offprints{meilland@mpifr-bonn.mpg.de}

\institute{Max Planck Intitut fur Radioastronomie, Auf dem Hugel 69, 53121 Bonn, Germany 
\and
Laboratoire A.H. FIZEAU, UMR 6525, UNS/OCA/CNRS, Campus Valrose, 06108 Nice cedex 2, France
\and
Laboratoire d'Astrophysique de Grenoble, UMR 5571 Universit\'e Joseph Fourier/CNRS, BP 53, 38041 Grenoble Cedex 9, France
\and
INAF-Osservatorio Astrofisico di Arcetri, Istituto Nazionale di Astrofisica, Largo E. Fermi 5, 50125 Firenze, Italy
}

   \date{Received; accepted}

   \abstract{}{We study the Be star $\delta$ Cen circumstellar
     disk using long-baseline interferometry which is the only
     observing technique capable of resolving spatially and spectroscopically  objects smaller than 5 mas in the H and K band.}
{We used the VLTI/AMBER instrument on January 6, 8, and 9, 2008, in the H and K bands to complete low (35) and medium (1500) spectral resolution observations.} 
{We detected an oscillation in the visibility curve plotted as a function of the spatial frequency which is a clear signature of a companion around $\delta$ Cen. Our best-fit soltution infers a binary separation of 68.7 mas, a companion flux contribution in the K band of about 7\% of the total flux, a PA of 117.5 $\degr$, and an envelope flux around the Be primary that contributes up to about 50 \% of the total flux, in agreement with our Spectral Energy Distribution (SED) fit. The envelope size is estimated to be 1.6 mas in K but no departure from spherical symmetry is detected. }{}

   \keywords{   Techniques: high angular resolution --
                Techniques: interferometric  --
                Stars: emission-line, Be  --
                Stars: winds, outflows --
                Stars: individual ($\delta$ Cen) --
                Stars: circumstellar matter
                Stars: binary 
               }

   \maketitle
%

\section{Introduction} \par

We have initiated an observational campaign to determine the global parameters of Be stars and the origin of the so-called ``Be-phenomenon" using the VLTI interferometer, which combines, for the first time, milli-arcsecond spatial resolution with good spectral resolution of up to 12000 in the near-IR K band. We are therefore able to study the geometry and kinematics within the circumstellar environment. Three Be stars have been observed with the VLTI instruments and the global physical properties of these stars have been found to vary strongly from one star to another.

Achernar was the first Be star observed with the VLTI using the VINCI beam recombiner. These observations exhibited a strongly flattened star due to an almost critical rotation (Domiciano de Souza \& Kervella 2003) and an extended polar wind that contributed to almost 4\% of the total flux in the H band (Kervella \& Domiciano de Souza 2007) with no detectable equatorial disk. Nevertheless, its circumstellar environment was known to be highly variable with a quasi-cyclic formation and a dissipation timescale of about 13 years proposed by Vinicius et al (2007). Kanaan et al. (2008) modeled these variations successfully as a possible outburst of circumstellar matter that has been ejected and was propagating within the stellar environment with an expansion velocity of 0.2km~s$^{-1}$.

$\alpha$ Arae was the second star observed with the VLTI, using both MIDI and AMBER instruments, and the observational results were published by Chesneau et al. (2005) and Meilland et al. (2007a). The data interpretation was completed with the SIMECA code developed by Stee et al. (1994), which is able to simulate the gaseous circumstellar environment of active hot stars. Two distinct regions in the circumstellar environment of  $\alpha$ Arae are clearly present : a thin Keplerian rotating equatorial disk and a rapidly expanding enhanced polar wind. As for Achernar, the stellar rotational velocity, which was readily measurable due to the flatness of the projected envelope onto the sky plane, was found to be almost critical. By combining MIDI and AMBER data, they found that the disk size did not vary as a function of the observing wavelength, which was a strong indication of possible disk truncation, as confirmed by more ``classical" spectroscopic measurements.

Finally, the third Be star studied using the VLTI instruments, was $\kappa$ CMa, observed with AMBER in MR mode and presented in Meilland et al. (2007b). For this star, a non-Keplerian rotating disk was detected. The star was found to rotate far below its critical velocity, which appears to indicate that critical rotation may not be a requirement of the ``Be phenomenon". Since it has an early spectral type of B2V, the radiation pressure might however be sufficient to initiate a mass loss and possible disk formation. This hypothesis agrees with the Cranmer (2005) statistical study. Finally, the spectro-differential interferometric measurements were able to detect the presence of an inhomogeneity in the circumstellar envelope of this star, which could be interpreted in terms of the "one-armed" viscous disk framework of Okazaki(~\cite{okazaki}).

To increase the number of Be stars observed by the VLTI and initiale a statistical study of the geometry and kinematics of their surroundings, and study the possible dependence of the ``Be phenomenon" on the spectral type of the central star, we requested new AMBER observations of 6 close and bright enough Be stars of spectral types ranging from B1 to B8 of which $\delta$ Cen was one star.

This letter is organized as follows: in Sect. 2 we briefly introduce the Be star $\delta$ Cen and its fundamental parameters, in Sect. 3 we present our VLTI/AMBER observations and the data reduction, and in the last section our results are summarized and conclusions are presented.

\begin{figure}[htbp]
\hspace{0.5cm}
{\begin{tabular}{ccc}
\hline  Name & L (m)& P.A. ($^o$)\\
\hline
\hline
B$_1$& 85 & -158 \\
B$_2$& 85.9 & -75  \\
B$_3$& 127.9 & -116\\
B$_4$& 31.9 & -111  \\
B$_5$& 16 & -111 \\
B$_6$& 47.8 & -111 \\
\hline
\end{tabular}}

	\vspace{-3.4cm}
	\hspace{4.2cm}
   \includegraphics[width=0.22\textwidth]{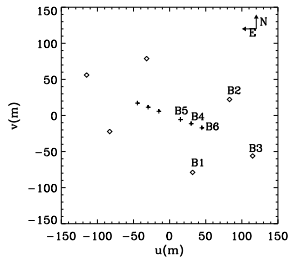}
\caption{Left : Names, lengths, and P.A. for each baseline of the two triplets of observation. Note that the two measurements from January 8 with A0-K0-G1 are merged into  B$_1$, B$_2$, and B$_3$ baselines, since the projected baseline length and P.A. does not vary significantly between them. B$_4$, B$_5$, and B$_6$ baselines correspond to the E0-G0-H0 triplets. Right: (u,v) plane coverage obtained during the $\delta$ Cen observing run. Fortunately, the largest projected baseline is aligned with the shorter ones.}
\label{all_visi}
\end{figure}

\vspace{-0.8cm}

\section{$\delta$ Cen}
$\delta$ Cen (HR 4621, HD 105435, HIP 59196) is usually classified as a B2IVne star. Levenhagen \& Leister (2006) considered a spectroscopic analysis that did not take account of the effects of gravity darkening and proposed a B2Vne classification. It is a variable star with a brightness variation between magnitude 2.65 and 2.51. $\delta$ Cen has the same proper motion as the nearby stars HD 105382 (which is also a Be star) and HD 105383 and may therefore be member of a triple visual star system. A precise radial velocity (RV) measurement campaign of B stars in the Scorpius-Centaurus association conducted by Jilinski et al. (2006) measured a RV for $\delta$ Cen of 3.8$\pm$ 2.8  km~s$^{-1}$. The RV determined by various authors from the SIMBAD database also exhibits a large dispersion, which might indicate possible binarity, as we see in the following. On the other side, Mason et al. (1996) and Mason (2008), using U.S. Naval Observatory (USNO) and Center for High Resolution Astronomy (CHARA) speckles cameras did not detect any binarity for this star.

Based on an autocorrelation analysis of Hipparcos photometry, Percy et al. (2002) presented a self-correlation diagram with strong minima at 0.55 and 1.1 days, in contrast, Balona (1990) reported a tentative period of 1.923 days, implying that $\delta$ Cen is a short-period variable Be star. Studying line profile variations in 1999 HEROS spectroscopic data, Rivinius et al. (2003), confirmed both values at 0.532d and 1.139d but with a formally somewhat higher significance of the longer period which was adopted in their paper. They concluded that $\delta$ Cen line profile variability was modeled successfully  as  a non-radial pulsator with $l$ = $m$ = +2, as for the Be star $\omega$~CMa, but with a strongly asymmetric power distribution of the line profile variability with a unidentified cause. The star's distance was estimated by Hipparcos to be 121$\pm$11pc, and its K and V magnitude to be 2.7 and 2.56, respectively . Lenorzer et al. (2002) identified $\delta$ Cen as a B star with emission lines observed in the ISO Short Wavelength Spectrometer  Post-Helium program of the Infrared Space observatory (ISO). The fundamental parameters determined for rapidly rotating B-type stars by Fr\'emat et al. (2005), taking into account gravitational darkening were T$_{eff}$=22384 $\pm$ 446 K, log $g$=3.942 $\pm$0.053, and Vsin$i$=263$\pm$14 km~s$^{-1}$. Its linear critical equatorial rotation velocity was V$_c$=527$\pm$29 km~s$^{-1}$ and its inclination angle estimated to be $i$=41.6$\pm$2.3$\degr$. Levenhagen \& Leister (2004) estimate its age to be log Age = 7.20 $\pm$0.15 yr, its luminosity log $L/L_{\sun}$=3.70 $\pm$0.10, and its mass 8.6 $\pm$0.3 M$_{\sun}$.

\section{Observations and data reduction}

\begin{table}
{\centering \begin{tabular}{cccccccc}
\hline  Obs. time &Target & Triplet & Mode & Seeing\\
\hline
\hline
06/01/08 08:59 & $\delta$ Cen & A0-K0-G1 & MR & 1.02"\\
06/01/08 09:11 & HD109538 & A0-K0-G1 & MR  & 1.13"\\
06/01/08 09:26 & $\delta$ Cen & A0-K0-G1 & MR & 1.19"\\
\hline
08/01/08 08:48 & HD90798 & E0-G0-H0 & LR & 0.50"\\
08/01/08 09:11 & $\delta$ Cen  & E0-G0-H0 & LR & 0.94"\\
\hline

09/01/08 08:07 & HD47536 & E0-G0-H0 & MR & 0.68"\\
09/01/08 08:31 & $\delta$ Cen & E0-G0-H0 & MR &  0.74" \\
09/01/08 08:50 & HD109538 & E0-G0-H0 & MR & 0.87"\\
09/01/08 09:08 & $\delta$ Cen & E0-G0-H0 & MR &  1.22"\\
09/01/08 09:28 & HD109538 & E0-G0-H0 & MR  & 0.71"\\
\hline
\end{tabular}\par}
\caption{Observations log of $\delta$ Cen and its calibrators during the January 2008 Be AMBER GTO run.}
\end{table}

$\delta$ Cen  was observed with AMBER in low and medium
spectral resolution modes (spectral resolving powers of R=35 and R=1500, respectively) during the
observing runs 080.C-0099 and 080.D-0131, completed in January, 6, 8, and 9, 2008. Five
measurments were obtained with the Auxilliary Telescopes (1.8m
diameter) on the A0-K0-G1 and E0-G0-H0 baseline triplets. HD~109593
(apparent diameter of 1.638$\pm$0.021\,mas in the CHARM2 catalog
from Richichi et al. \cite{2005A&A...431..773R}), HD~47536 (apparent diameter of
1.69$\pm$0.03\,mas, also in CHARM2), and HD~90798 (apparent diameter
1.70$\pm$0.51\,mas using the getCal\footnote{available at
  \url{http://mscweb.ipac.caltech.edu/gcWeb}} tool) were
also observed to calibrate $\delta$ Cen visibilities. The log of
these observations are presented Table 1.

The data were reduced using the AMBER data reduction software
developped by the JMMC: \texttt{amdlib}, version 2.1. It computes a fringe
fitting of the AMBER data, after correcting for detector
bias and flat-field (Millour et al., \cite{2004SPIE.5491.1222M} and
Tatulli et al., \cite{2007A&A...464...29T}). Following that step, we 
adopted two different approaches to calibrate the data: 
\begin{itemize}
\item The ``standard'' approach, which retains 20\% of the total number
  of frames according to the fringe SNR. This allows us to limit the
  fringe smearing effect (also called ``jitter'' effect) and the
  coherence length visibility loss due to the atmospheric OPD
  (described in Millour et al. \cite{2007MillourESOCalWS}). The
  following steps (transfer function estimation, calibration) were
  completed using a new set of custom-made scripts described by Millour et
  al. (\cite{MillourSpieDRS2008}\footnote{The scripts are available to
    the community at the following webpage:
    \url{http://www.mpifr-bonn.mpg.de/staff/fmillour}}).
\item An approach using bootstrapping, roughly explained in Tatulli
  et al., (\cite{2007A&A...464...29T}) and described in more detail in
  Cruzal\`ebes et al. (\cite{2007CruzalebesESOCalWS}). A frame-selection scheme
  that rejects frames with a negative flux, a high OPD variation from
  frame to frame, and a SNR of less than 1, is also applied. The calibration procedure is then
  completed the same way as for the previous method.
\end{itemize}

The two methods provide the same results within the error bars and
we therefore choose to use the calibrated data from the first method later in the paper. Since the aim of the paper is to determine the binary parameters, we exclude the January 9 MR data for which
projected baselines are similar to the January 8 LR ones and the
data quality is significantly lower. Finally, we merged the two
January 5 MR measurements obtained with almost the same projected
baseline, and determined the wavelength-dependent visibilities
for the six baselines presented Fig. 1. The corresponding (u,v)
plane coverage of the science target is also plotted in Fig 1.

\section{Results}

{\begin{figure}[t]
       \centering  \includegraphics[width=0.4\textwidth,height=0.19\textheight]{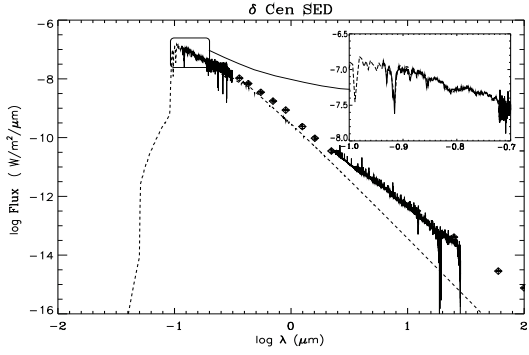}
\caption{$\delta$ Cen Spectral Energy Distribution (SED) reconstructed from various sources in the literature : Morel (1978), 2MASS, IRAS magnitude and flux measurements (diamonds), IUE spectra (solid line in the UV), and ISO spectra (solid line in the IR). The dashed line is a Kurucz model for T$_{eff}$~=~21000K, R$_\star$~=~5.9R$_\odot$, log~g~=~4, and d~=~121pc.}
\label{SED}
\end{figure}
\par}

{
\begin{figure*}[htbp]
       \centering \includegraphics[height=0.225\textwidth]{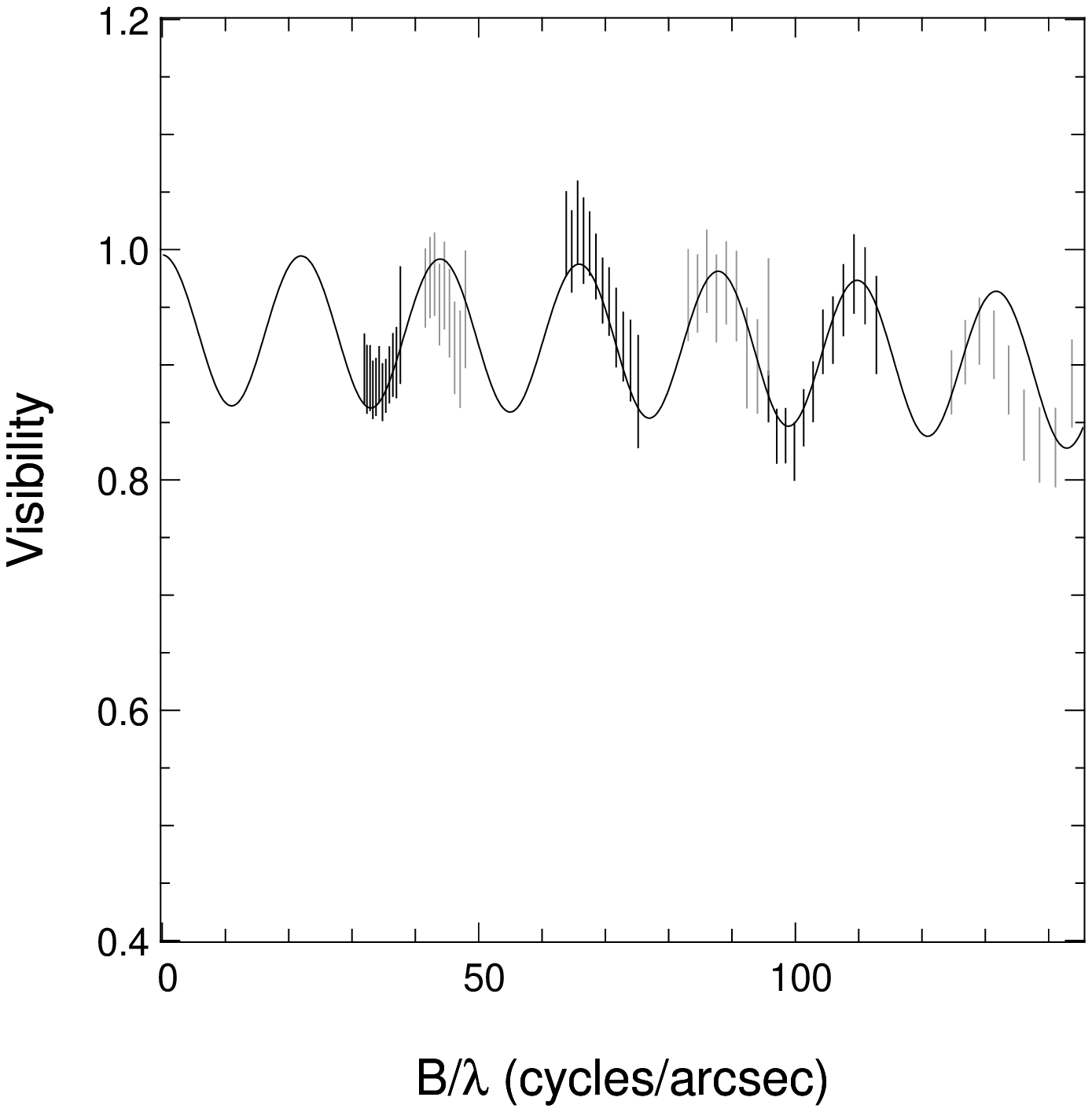}
       \centering \includegraphics[height=0.225\textwidth]{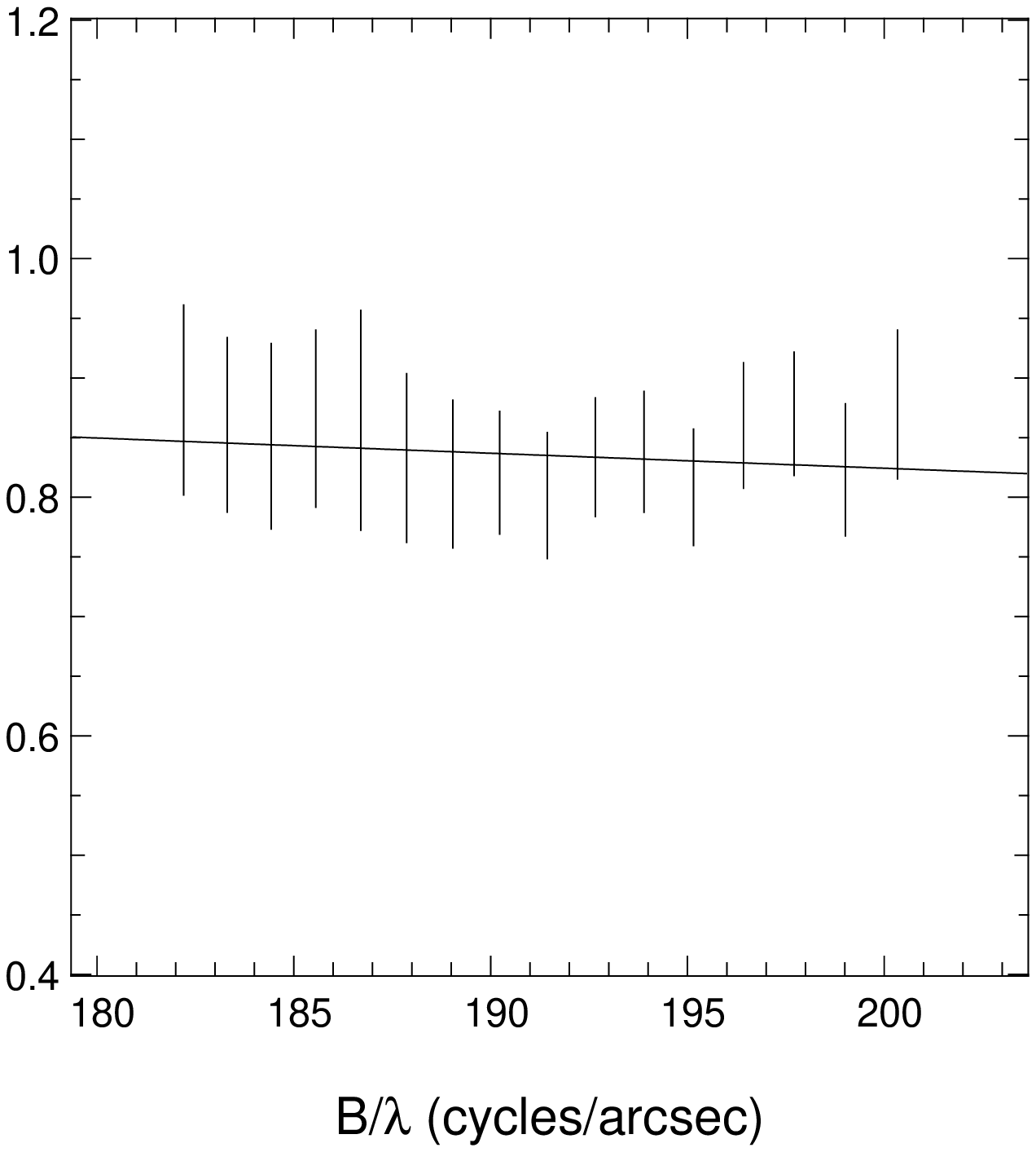}
       \centering \includegraphics[height=0.225\textwidth]{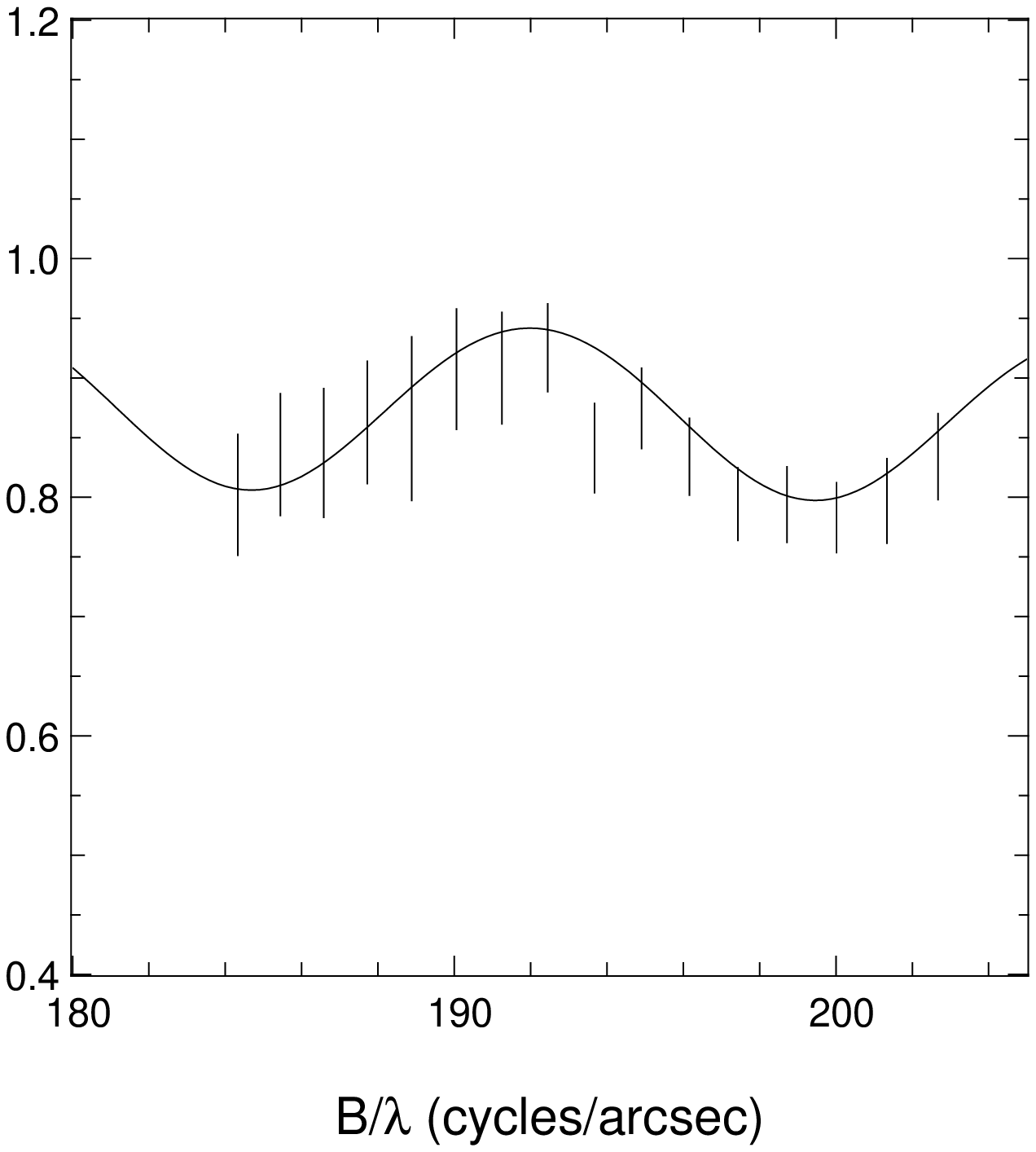}
       \centering \includegraphics[height=0.225\textwidth]{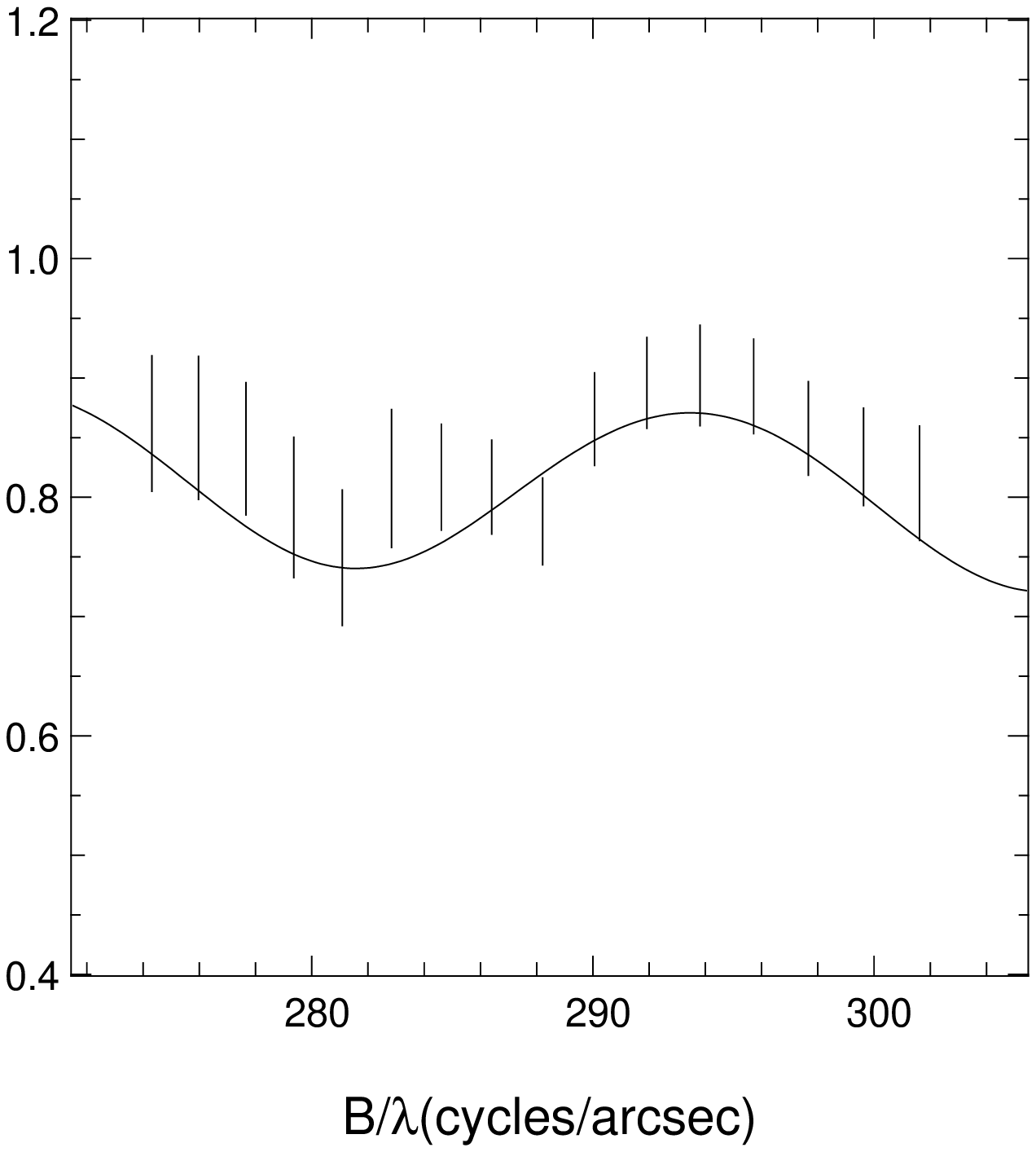}

\caption{Left: LR K (dark) and H (gray) bands visibilities for the January, 8 2008 baselines (B$_4$, B$_5$, and B$_6$) as a function of the spatial frequency. Second picture from the left to the right: MR visibilities for the January, 5 2008 baselines (B$_1$, B$_2$, and B$_3$) as a function of the spatial frequency. For all figures, measurements are given by the error bars, whereas the solid line corresponds to our best-fit binary + uniform disk circumstellar envelope modeled using parameters from Table 2.}

\label{all_visi}
\end{figure*}
\par}

For a single Be star, assuming that the visibility we measure, $V$, is only due to the central star and its circumstellar disk we can write: 

\begin{equation}
V= \frac {V_{env} F_{env} + V_{\star} F_{\star}} {F_{tot}}
\label{Eq.1}
\end{equation}
where V$_{env}$ and F$_{env}$ are respectively the envelope visibility and flux in the continuum , V$_{\star}$ and F$_{\star}$ the star visibility and flux in the continuum and F$_{tot}$ = F$_{env}$+F$_{\star}$. To measure accurately the extension of a circumstellar envelope from an interferometric measurement, we therefore need to determine both F$_{\star}$ and F$_{env}$ at the observing wavelength. This can be achieved directly by analyzing data acquired by interferometry if, at least, two baselines resolve the envelope completely but not the star (i.e, V$_{env}$=0 and V$_{\star}\sim$1). In the case of our $\delta$ Cen AMBER data, even the longest baseline (i.e, B$_3$=127.9m) is too short to fully resolve the circumstellar envelope. 

To determine F$_{\star}$ and F$_{env}$ in H and K bands to the highest possible accuracy, we started our modeling by constructing the $\delta$~Cen Spectal Energy Distribution (SED) using various sources in the literature (see Fig. 2). A Kurucz model, fitted from UV and visible measurements, for which the circumstellar contribution was negligible, was used to determine the stellar contribution to the flux for wavelengths ranging from UV to Far-IR. The values of the best-fit Kurucz model parameters, which were consistent with those found in the literature and were introduced in Sect. 1, are presented Table 3. This method allows us to characterize precisely the circumstellar IR-excess. Taking into account the uncertainty in the spectral class of $\delta$ Cen and its small variability in the near IR, we obtain F$_{env}$= 0.52$\pm$0.06~F$_{tot}$ in the K band and 0.38$\pm$0.08~F$_{tot}$ in the H band.

K and H band calibrated visibilities obtained with the small aligned triplet in LR mode (i.e. B$_4$, B$_5$, and B$_6$) are plotted as a function of the spatial frequency in Fig. 3. These data exhibit clearly a sinusoidal modulation with a period of T$_{4,5,6}$=22 cycles/arcsec and an amplitude of A$_{4,5,6}$=0.15$\pm$0.02. This variation cannot be reproduced by a simple star +  circumstellar envelope model regardless of the geometry of this second component (e.g. uniform or Gaussian disk, or ring), and is usually produced by a companion. The separation between the two components in the direction of the projected baselines (i.e., -111$^o$) is given by S$_{4,5,6}$=1/T$_0$=44.8$\pm$0.5 mas, and its flux  by F$_c$=A$_{4,5,6}$/2 = 0.075$\pm$0.01F$_{tot}$. We note that this value may be biased towards a dimmer flux ratio since the separation is $\sim$25$\%$ of the telescopes PSF ($\sim$~240mas with the ATs at 2.1$\mu$m) and the AMBER instrument uses monomode fibers (Tatulli et al. 2004). However, it is beyond the scope of this paper to correct for this effect.

To reproduce the small slope in the visibilty as a function of spatial frequency, we need to introduce a small circumstellar envelope (i.e., D=2$\pm$0.5mas) around one of the two components. We note that the data also indicate some small differences in modulation amplitude between the H and K band data. The characterization of these differences, probably due to the variation in the binary flux ratio between these spectral bands, is again beyond the scope of this letter and is not described here.

The B$_1$, B$_2$, and B$_3$ data are also plotted in Fig 3. These measurements place additional constraints on the binary and the circumstellar envelope. The B$_2$ and B$_3$ data show a sinusoidal modulation with periods corresponding to a separation between the binary components in the projected baseline directions of S$_2$=62$\pm$5~mas and S$_3$=52$\pm$6~mas, respectively. The amplitude of the modulation is A$_2$=A$_3$=0.15$\pm$0.05 which is in agreement with the measurement for the January 8 LR data (i.e., A$_0$=0.15$\pm$0.02). The separation measured by B$_3$ is consistent with that measured for data acquired with the LR triplet; this is a good indication that our measurements are not biased by some instrumental effects since B$_3$ is almost aligned with the LR triplet (i.e. P.A.=-116$^o$ for B$_3$, and P.A.=-111$^o$ for B$_4$, B$_5$, and B$_6$). The difference between the two measurements can be due either to a inaccurate spectral calibration of AMBER LR data (up to 0.1$\mu$m) or to a small change in the projected separation due to the 5$^o$ of variation in P.A. between the two data sets. Finally, there is no evidence of a sinusoidal modulation in the B$_1$ data. Since the modulation may not be detected if its period is at least 3 times the observation range in spatial frequency (i.e., T$\geq$50 cycles/arcsec), the binary separation along this projected angle, S$_1$, is lower than 20~mas. The average visibility of each baseline (i.e. $\overline{V}_1$=0.82, $\overline{V}_2$=0.86, and $\overline{V}_3$=0.81) indicates clearly the presence of an extended component, probably a circumstellar envelope with an extension of about 2$\pm$0.5mas.

By measuring the projected separation in at least two different directions in the sky, we can determine the modulus and the orientation of the binary separation on the sky plane. Our most appropriate binary star + circumstellar envelope model is obtained for a set of values given in Table 2. The modeled visibility obtained in K band for each of the six baselines is plotted with the observed measurements in Fig. 3. We note that to verify that the binary signal did not originate in the calibrators we checked that the modulation was evident in the science object raw visibilities and not on the calibrator ones.

Since AMBER operates simultaneously with three telescopes, we are able to measure the closure phase for each triplet. Such measurements place additional constraints on our models. Figure 4 shows a reasonable agreement between observed and modeled closure phase using parameters from Table 2.

Finally, using R$_\star$=5.9R$_\odot$ and d=121pc, the 68.7mas separation between $\delta$ Cen and its companion corresponds to 300$\pm$50R$_\star$. Assuming a circular orbit, this provides a lower limit to the semi-major axis. The companion spectral class can be inferred using both $\delta$ Cen spectral class and K band flux ratio determined from the amplitude of the modulation in the interferometric data. The companion spectral class should range between B4V and A0III, and its mass between 4 and 7 M$_\odot$. Using Kepler's third law of planetary motion, $P=\sqrt{a^3/M_{tot}}$, where $P$ is the period in years, $a$ is the semi-major axis in AU, and M$_{tot}$ is the binary system total mass in $M_\odot$, we can determine a lower limit to its period, i.e. 4.6 years with $M_{tot}$=15.2M$_\odot$ and $a$=250R$_\star$=6.9AU.

Additional interferometric observations, including long-term monitoring, would be required to determine why these results disagree with those of the speckles observations of Mason (1997, 2008) and to fully determine the projected orbit of the companion of $\delta$ Cen. Spectroscopic follow-up should also be completed even though the Doppler shifts due to binarity are difficult to measure if the star is a rapid rotator (i.e, V~sin~i= 263km s$^{-1}$), and the profiles are affected by non-radial pulsations. The circumstellar environment geometry and kinematics will be studied in a forthcoming paper, which will include all January 2008 VLTI/AMBER medium resolution differential visibilties as well as VLTI/MIDI data. 

\vspace{-0.4cm}

\begin{table}[h]

\begin{tabular}{cc}
\hline
Distance & 121 $\pm$ 11 pc\\
$T_\mathrm{eff}$& 21000 $\pm$ 1000 K\\
Stellar Radius & 5.9 $\pm$0.5 R\( _{\sun } \)\\
log $g$ & 3.75 $\pm$ 0.25\\
\hline
\hline
Companion flux (K band) & 0.07$\pm$0.0.1 F$_{tot}$\\
Separation & 68.7$\pm$0.5 mas\\
Position Angle (PA) & 117.5$\pm$0.5$\degr$\\
Envelope flux (K band) & 0.52$\pm$0.06  F$_{tot}$\\
Envelope diameter (K band) & 1.6$\pm$0.4 mas\\
\hline
\end{tabular}
\vspace{0.2cm}
\centering \caption{Parameters and results for the binary system + uniform disk circumstellar envelope obtained for $\delta$ Cen. Note that there is a 180$^o$ ambiguity in the position angle since the AMBER closure phase data reduction process is not fully validated.}
\end{table}

\begin{figure}[htbp]
       \centering  \includegraphics[width=0.225\textwidth]{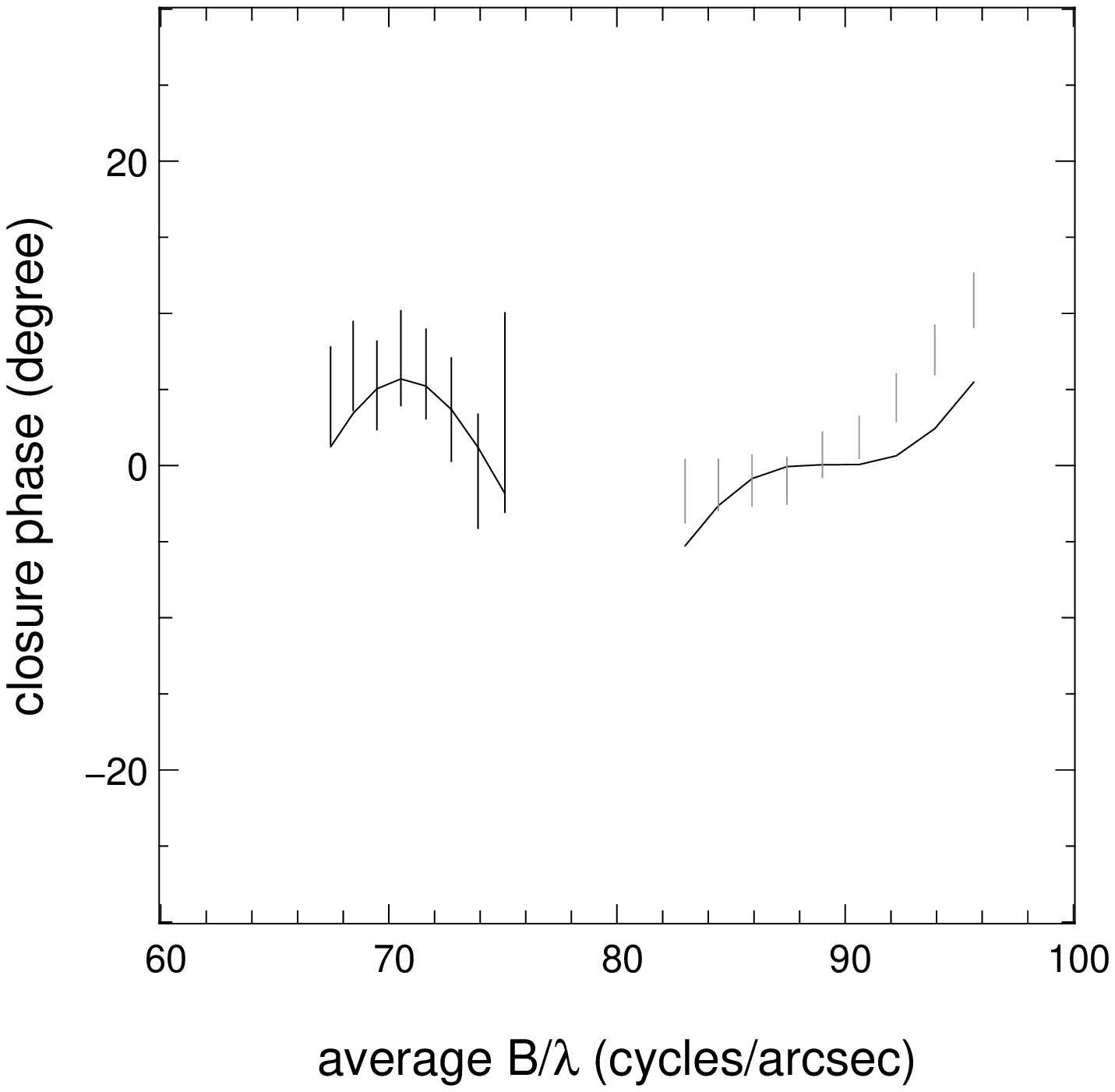}
       \centering  \includegraphics[width=0.225\textwidth]{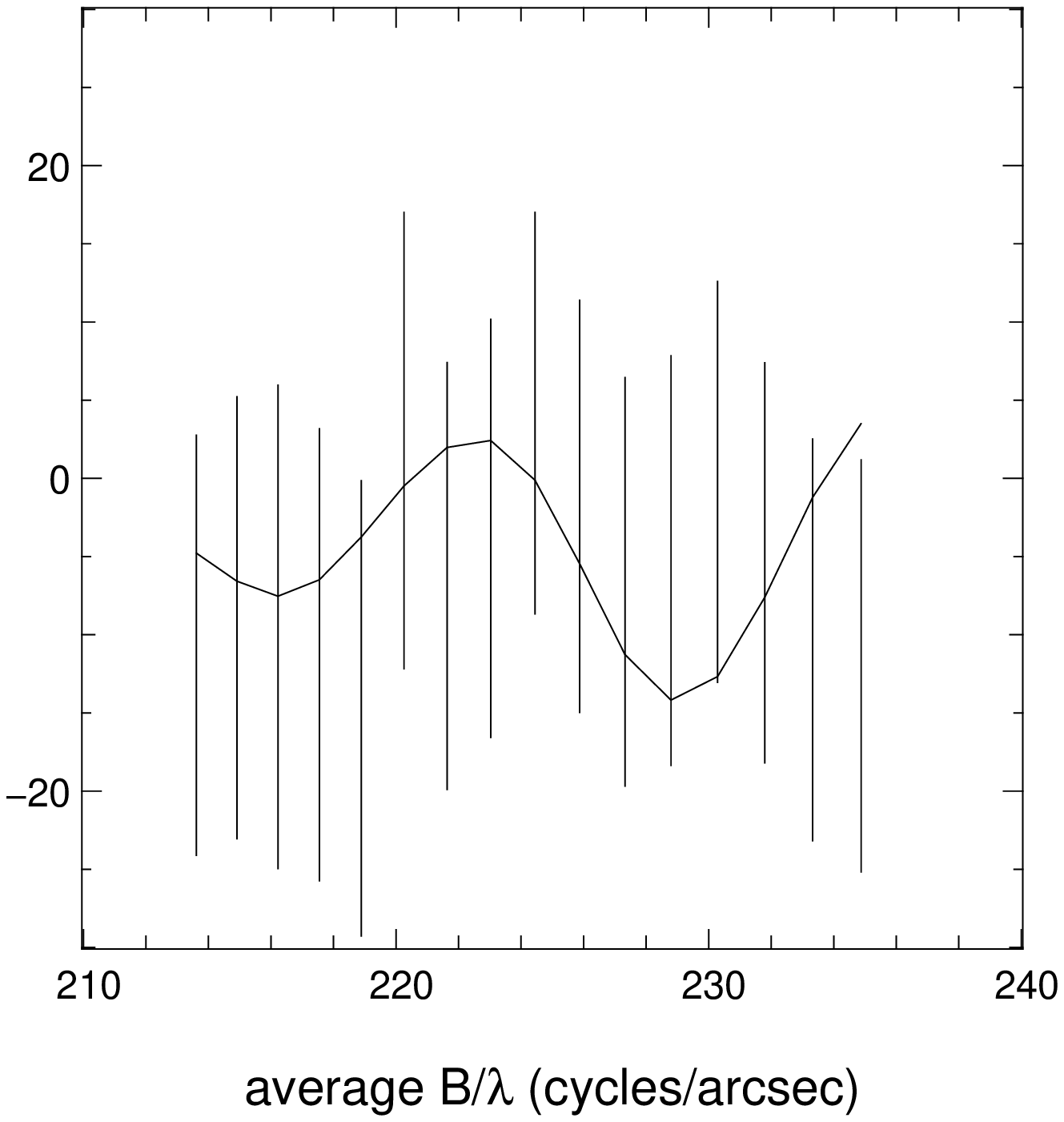}
\caption{Observed and modeled closure phase for the B$_4$B$_5$B$_6$ (left) and B$_1$B$_2$B$_3$ (right) triplets plotted as a function of the average spatial frequency. For all figures, the measurements are represented by the error bars whereas the solid line corresponds to our best binary + uniform disk circumstellar envelope modeled using parameters from Table 2.
}
\label{cp}
\end{figure}

\vspace{-1cm}

\begin{acknowledgements}
The Programme National de Physique Stellaire (PNPS), the Institut National en Sciences de l'Univers (INSU)  and the Max Planck Institut fur Radioastronomy (MPIfr) are acknowledged for their financial support. 
The authors would like to thank ESO staff for their help to make the observations a success and more particularly J-B Lebouquin and F. Rantakyro. This research has made use of SIMBAD database, operated at CDS, Strasbourg, France.
\end{acknowledgements}

\end{document}